\documentclass[3p, twocolumn]{elsarticle}

\usepackage{amssymb}
\usepackage{amsmath}
\usepackage{epsfig}
\newcommand{\BG}{\rm\scriptscriptstyle{BG}}
\begin{document}


\title {Anomalous electrical conductivity in rapidly crystallized~ Cu${}_{50-x}$Zr${}_{x}$ (x = 50 - 66.6) alloys}

\author[imet]{S.A.~Uporov\corref{cor1}}
\author[imet]{S.~Estemirova}
\author[CSUN]{N.M.~Chtchelkatchev}
\author[imet,urfu]{R.E.~Ryltsev}
\address[imet]{Institute of Metallurgy, Ural Division of Russian Academy of Sciences, Amudsena str. 101, 620016 Ekaterinburg, Russia}
\address[urfu]{Ural Federal University named after First President of Russia B.N. Yeltsin, Mira str. 19, 620002 Ekaterinburg, Russia}
\address[CSUN]{Department of Physics and Astronomy, California State University Northridge, Northridge, CA 91330}
\cortext[cor1]{Corresponding author. Tel.: +79089192999
 E-mail: segga@bk.ru}

\begin{abstract}
Cu${}_{50-x}$Zr${}_{x}$ (x = 50, 54, 60 and 66.6) polycrystalline alloys were prepared by arc-melting. The crystal structure of the ingots has been examined by X-ray diffraction. Non-equilibrium martensitic phases with monoclinic structure were detected in all the alloys except Cu${}_{33.4}$Zr${}_{66.6}$. Temperature dependencies of electrical resistivity in the temperature range of T = 4 - 300 K have been measured as well as room temperature values of Hall coefficients and thermal conductivity. Electrical resistivity demonstrates anomalous behavior. At the temperatures lower than 20 K, their temperature dependencies are non-monotonous with pronounced minima. At elevated temperatures they have sufficiently non-linear character which cannot be described within framework of the standard Bloch--Gr\"{u}neisen model. We propose generalized  Bloch--Gr\"{u}neisen model with variable Debye temperature which describes experimental resistivity dependencies with high accuracy. We found that both the electrical resistivity and the Hall coefficients reveal metallic-type conductivity in the Cu-Zr alloys. The estimated values of both the charge carrier mobility and the phonon contribution to thermal and electric conductivity indicate the strong lattice defects and structure disorder.
\end{abstract}


\maketitle

\section{Introduction}

The Cu-Zr system is one of most extensively studied ones among binary metal-metal glass-forming alloys. These alloys attract attention due to their ability to form bulk metallic glasses \cite{Xu,Wang1,Wang2}. It should be noted that the best glass formers in the system are located at so-called pinpoint compositions \cite{Wang3,Shen,Wang4,Yang1}. The bulk amorphous ingots can be prepared only at these narrow concentration intervals.  Large amount of research in the Cu-Zr system is devoted to investigation of amorphous phase formation and crystallization kinetics. Recently a number of  criteria to estimate glass-forming compositions have been proposed \cite{Yang1,Yu,Bendert,Yang2,Kwon}. However, the physical nature of good glass formation at the pinpoints is still unclear. To study these issues, the alloys in amorphous state are usually considered while physical properties of crystalline Cu-Zr alloys have not been systematically investigated. There are only few reports devoted to this problem \cite{Gantmakher1,Gantmakher2,Carvalho,Garoche,Glimois}. In our opinion, investigation of the Cu-Zr alloys in crystalline state may also give useful information to understand the nature of their glass-forming ability.

A study of transport properties, such as electrical and thermal conductivity and Hall coefficient, gives information on charge scattering mechanisms. Quenched Cu-Zr alloys tend to form metastable and disordered phases, such as martinsitic monoclinic structures \cite{Fornell}. Obviously, any structural defects affect electron transport noticeably. Authors of \cite{Gantmakher1} have shown that the electrical resistivity of some Cu-Zr alloys demonstrates significantly nonlinear temperature dependences in crystalline state. Besides, positive Hall coefficients and large electrical resistivity (100 - 150 $\mu $Ohm$\times$cm) have been observed for the alloys at room temperature. These results have been interpreted within the framework of two-band model which implies high density of $d$-states at the Fermi level and takes into account the Mott $s-d$ interband scattering \cite{Mott}. However, the electronic structure of the Cu-Zr intermetallic compounds investigated by first principles calculations \cite{Hasegawa,Du} demonstrates low electron density of states at the Fermi energy. It has been suggested by J. Du et al. \cite{Du} that the CuZr${}_{2}$ compound is semiconductor with indirect band gap of 0.227 eV, while the others intermetallics are conductors.

Thus the electronic transport properties of crystalline Cu-Zr alloys and their relation with crystal structure are still unclear. This work is a milestone on this way.  We study crystal structure, electrical resistivity, Hall coefficients and thermal conductivity of arc melted polycrystalline Cu${}_{50-x}$Zr${}_{x}$ (x = 50, 54, 60 and 66.6) alloys and observe non-trivial behavior of these properties.

\section{Experiment}

Zirconium and copper with purity 99.98 mass. \% were used to prepare the Cu-Zr alloys with nominal compositions of Cu${}_{50}$Zr${}_{50}$, Cu${}_{46}$Zr${}_{56}$, Cu${}_{40}$Zr${}_{60}$ and Cu${}_{33.4}$Zr${}_{66.6}$. The polycrystalline alloys have been obtained using standard arc-melting technique under helium atmosphere. The samples were re-melted at least five times to ensure their homogeneity. As-cast alloys have been obtained by quenching of the melts on furnace mold with a cooling rate of about 100 K/sec.

The crystal structure and phase content of the samples were studied using X-ray diffraction analysis (XRD) with Shimadzu XRD-7000 diffractometer. The XRD patterns were obtained using CuK${}_{\alpha}$ - radiation, graphite monochromator, the 2$\Theta$ range of 25-100 deg, scan step of 0.04º and scan exposure of 3 second / step. The unit cell parameters were calculated using the RTP software \cite{Chebotarev}.

The compositions of the alloys as well as the concentrations of impurities were controlled by atomic-emission method using the SpectroFlame Modula S analyzer.

Measurements of electrical resistivity and Hall coefficients of the samples were carried out using Cryogenic VSM CFS-9T-CVTI system and standard four-probe method. In course of the conductivity and Hall measurements we used Keithley K2400 as a source of direct current (DC), and Keithley K2182 nanovoltmeter for recording the voltage data. The magnitude of DC was chosen to be 100 mA. The samples for the resistivity investigations were prepared in a rectangular form with sizes of length $\times$ width $\times$ height, 8 mm $\times$ 3 mm $\times$ 3 mm, respectively. Soldering by pure indium was used to achieve the electrical contacts between the samples and holder. The electrical resistivity data as a function of temperature were collected during continuous cooling with rate of 0.7 K/min in the temperature range of 4 -- 300 K. The transverse voltage as a function of magnetic field was determined to obtain the Hall coefficients of the alloys at 300 K. At the same temperature, we calculated thermal conductivity of the samples as a product of thermal diffusivity, specific heat and density. The thermal characteristics (thermal diffusivity, specific heat) were determined by laser flash method (LFA) using Netzsch LFA 457 device. The hydrostatic weighing was used to measure density of the alloys.

\section{Results and discussion}

Fig.~1 shows XRD patterns of as-cast Cu-Zr alloys. Analysis of XRD data reveals that  Cu${}_{50}$Zr${}_{50}$ alloy consists of two monoclinic phases corresponding to the basic structure (P2${}_{1}$/m space group) and the superstructure (Cm space group) \cite{Carvalho,Fornell,Schryvers}. It is well known that rapidly quenched CuZr${}_{ }$ intermetallic compound undergoes martensitic transformation from B2 into two monoclinic structures of the same symmetries \cite{Schryvers}. According to equilibrium phase diagram \cite{Okamoto}, the B2 phase decomposes into two phases Cu${}_{10}$Zr${}_{7}$ and CuZr${}_{2}$ with orthorhombic and tetragonal structure respectively. So we conclude that the structure of as cast Cu${}_{50}$Zr${}_{50 }$ alloy contains of metastable martensitic phases only.

\begin{figure}[t]
  \centering
  \includegraphics[width=\columnwidth]{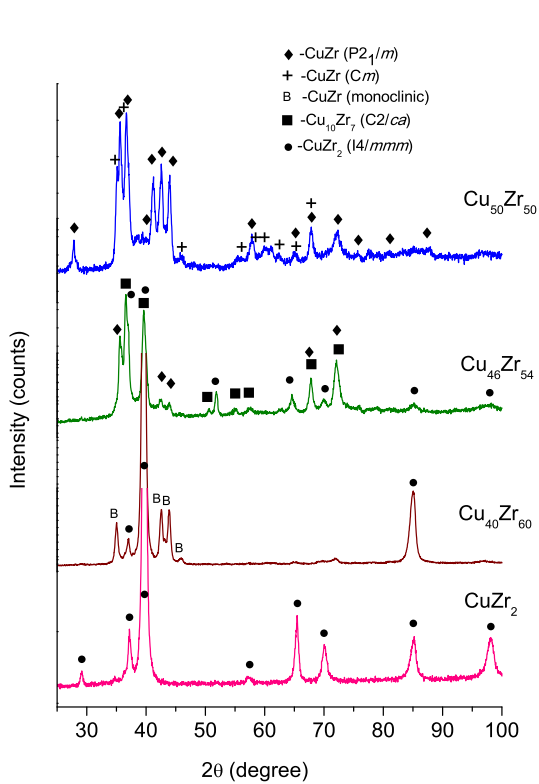}\\
  \caption{(Color online) XRD patterns of Cu-Zr alloys.}
  \label{fig1:XRD}
\end{figure}

\begin{table*}
\centering
\scriptsize
\begin{tabular}{|c|c|c|c|c|c|c|} \hline
Alloy & Phases / fraction & \multicolumn{5}{|p{2.8in}|}{lattice parameters} \\ \hline
 &  & a, {\AA} & b, {\AA} & c, {\AA} & $\beta$,$\circ$ & V, ${{\AA}}^{3}$ \\
 Cu${}_{50}$Zr${}_{50}$ & monoclinic CuZr \textit{P2${}_{1}$/m } (71 \%) & 3.310(4) & 4.1139(4) & 5.337(7) & 102.7(1) & 71.3(3) \\
 & monoclinic CuZr \textit{Cm} (29 \%) & 6.417(7) & 8.535(8) & 5.743(9) & 108.1(1) & 299(1) \\ \hline
  & monoclinic CuZr \textit{P2${}_{1}$/m} (5 \%) & 3.32(1) & 4.157(3) & 5.15(2) & 104.2(7) & 68.9(7) \\
Cu${}_{46}$Zr${}_{54}$ & tetragonal CuZr${}_{2}$ \textit{I4mmm} (50 \%) & 3.2298(7) &  & 11.25(1) &  & 117.4(2) \\
 & orthorhombic Cu${}_{10}$Zr${}_{7}$ \textit{Ñ2/ña} (45 \%) & 12.67(2) & 9.338(6) & 9.349(5) &  & 1106(3) \\ \hline
 Cu${}_{40}$Zr${}_{60}$ & tetragonal CuZr${}_{2}$ \textit{I4mmm} (92 \%) & 3.2227(9) &  & 11.163(9) &  & 115.9(2) \\
 & monoclinic CuZr  (8 \%) & 5.092(8) & 2.658(6) & 5.234(9) & 100.5(1) & 69.6(4)\\ \hline
  CuZr${}_{2}$ & tetragonal CuZr${}_{2}$ \textit{I4mmm} (100 \%) & 3.2303(9) & - & 11.10(4) & - & 115.8(4) \\ \hline
\end{tabular}
\caption{Structural parameters of the Cu-Zr alloys.}\label{Tab1}
\end{table*}

The XRD data for Cu${}_{46}$Zr${}_{54}$ indicate the presence of equilibrium phases CuZr${}_{2}$ and Cu${}_{10}$Zr${}_{7}$, and also a small amount of  P2${}_{1}$/m martensitic CuZr phase. In Cu${}_{40}$Zr${}_{60}$ alloy, the equilibrium CuZr${}_{2}$ phase and primitive monoclinic structure were revealed. The Cu${}_{33.4}$Zr${}_{66.6}$ alloy has the only CuZr${}_{2 }$ phase with tetragonal I4/mmm crystal structure.

 Note that the monoclinic structure observed in Cu${}_{40}$Zr${}_{60}$ alloy is rather unusual one. As far as we know, there is the only reference to this structure in Cu-Zr system \cite{Zhalko-Titarenko} where the authors recognized it as the second martensitic phase in rapidly cooled CuZr compound (instead of Cm phase usually observed in this case). We suggest that the phase under consideration is one of metastable martensitic structures which can form at certain condition. As we will show below, the existence of this phase affects the electronic structure and transport properties of the system.

 As seen in the Fig.~1, Bragg reflections in XRD patterns of the alloys are broad enough. The half width at half maximum of the diffraction peaks is estimated to be in the range of 0.5 - 1 deg. It can be explained by strong structure disorder due to high density of lattice defects. Such structure is expectable for the rapidly cooled alloy with competition of different structures, especially the martensitic ones. The lattice parameters of the detected phases as well as their fractions are represented in Tab.~1. The phase content has been estimated with accuracy of $\pm$ 2 vol. \%.

\begin{figure}[t]
  \centering
  \includegraphics[width=\columnwidth]{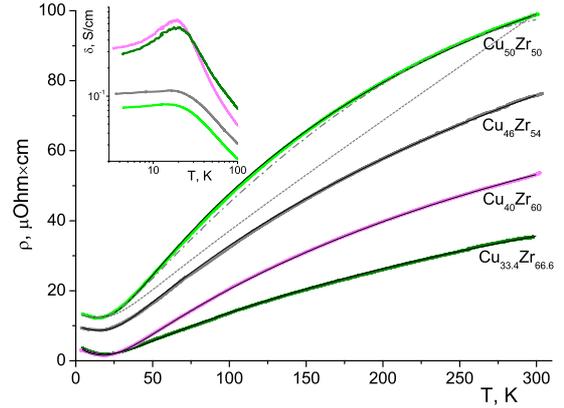}\\
  \caption{(Color online) The temperature dependencies of electrical resistivity of Cu-Zr alloys at different compositions. Thin solid lines represent the best fit by formula~\eqref{resist} with the modified Bloch--Grüniesen relation discussed in text. The dashed and dot-dashed lines for Cu${}_{50}$Zr${}_{50 }$ alloy represent the best fits of experimental data by formula~\eqref{resist} and the one with using the additional Mott term respectively. The inset shows the enlarged fragment of the low-temperature electrical conductivity.}
  \label{fig2:XRD}
\end{figure}

The temperature dependence of electrical resistivity $\rho $ of the Cu-Zr alloys is presented in Fig.~2. We see that the $\rho (T)$ dependencies have a non-linear behavior with a positive slope in the temperature range of 20 - 300 K. At temperatures lower than 20 K, non-monotonous behavior of resistivity with pronounced minima is observed for all the samples. The inset in Fig.~2 shows the enlarged fragments of the low-temperature conductivity curves which has Kondo-like type and vary their shapes with zirconium content.

 According to the results of chemical analysis, the initial zirconium and copper as well as prepared samples contain uncontrollable impurities of iron (up to 200 ppmw), cobalt (up to 20 ppmw), nickel (up to 450 ppmw) and manganese (up to 20 ppmw). So we suggest that abnormal low-temperature behavior of resistivity may be manifestation of the Kondo-effect caused by interaction of magnetic impurity atoms with nonmagnetic atoms of the alloy. The resistivity minimum temperatures \textit{T${}_{K}$}, so-called Kondo ones, are collected in the Tab.~2. To estimate amplitude of the observed effect, we calculate the relative increase of resistivity using the ratio:
$$
\Delta =\frac{\rho (4K)-\rho (T_{K} )}{\rho (4K)} \times 100\%
$$
where $\rho(4K)$, \textit{$\rho $(T${}_{K}$)} are resistivities at 4 K and T${}_{K}$ temperatures respectively. These values are also listed in Tab. 2.

As seen in Fig.~2, the temperature dependencies of electrical resistivity demonstrate metallic character above 20 K for all the compositions investigated. Note that for all the temperatures studied,  the absolute values of the resistivity decrease with increase of zirconium content and so minimal resistivities were observed in Cu${}_{33.4}$Zr${}_{66.6}$ alloy which is in fact CuZr${}_{2 }$ intermetallic compound. Moreover, the values of resistivity of this compound are lower than those for pure zirconium. These observations are in sharp contrast with usual behavior of resistivity of metallic alloys \cite{Bauccio_book} and also contradicts with theoretical predictions made in \cite{Du} where CuZr${}_{2 }$ is considered to be a semiconductor with pronounced pseudo-gap at Fermi level.

\begin{table*}
\centering
\scriptsize
\begin{tabular}{|c|c|c|c|c|} \hline
& Cu${}_{50}$Zr${}_{50}$ & Cu${}_{46}$Zr${}_{54}$ & Cu${}_{40}$Zr${}_{60}$ & Cu${}_{33.4}$Zr${}_{66.6}$ \\ \hline
T${}_{K}$, K \newline ($\pm$ 0.5 K) & 14.0 & 14.5 & 18.5 & 19.5 \\ \hline
$\Delta$, \% & 5.6 & 8.5 & 97.8 & 88.4 \\ \hline
$\rho $(300 K),\newline $\mu $Ohm$\times$cm & 98.8 & 76.1 & 53.1 & 35.4 \\ \hline
$\rho $(4 K),\newline $\mu $Ohm$\times$cm & 13.2 & 9.5 & 3.3 & 3.7 \\ \hline
$\rho $${}_{0}$, $\mu $Ohm$\times$cm & 11.0 & 8.0 & 1.0 & 1.2 \\ \hline
$\Theta $${}_{0}$, K & 170 & 165 & 165 & 150 \\ \hline
$\alpha $ & 0.20 & 0.12 & 0.14 & 0.11 \\ \hline
D,  & 70 & 30 & 25 & 25 \\ \hline
$\mu $, 10${}^{8}$ & 0.75 & 0.75 & 1.35 & 1.40 \\ \hline
\end{tabular}
\caption{The values of T${}_{K}$, $\Delta$, $\rho $ and the fitting parameters ($\rho $${}_{0}$, $\Theta $${}_{0}$, $\alpha $, D and $\mu $) for the Cu-Zr alloys.}\label{Tab2}
\end{table*}

As was mentioned above, the electrical resistivity of the studied Cu-Zr alloys exhibits  metallic behavior and low-temperature Kondo-like anomaly. So it is natural to fit experimental $\rho (T)$ curves by the combination of the Bloch--Gr\"{u}neisen relation $\rho _{\BG} (T)$ describing the contribution to resistivity due to electron--phonon scattering \cite{Abrikosov} and the Kondo term $\rho _{K} (T)$ \cite{Abrikosov, Kondo}:

\begin{multline}\label{resist}
\rho (T)=\rho_{0} +\rho _{\BG} (T)+\rho _{K} (T)=
\\
=\rho_{0} +A\frac{T^{5} }{\Theta ^{6} } \int _{0}^{\Theta /T}\frac{x^{5} }{(1-e^{-x} )(e^{x} -1)} dx +D\ln \frac{\mu }{T}.
\end{multline}
 Here \textit{$\rho $${}_{0}$} is the residual resistance due to static defects in the crystal lattice and uncontrolled impurities; \textit{A} is the electron--phonon coupling constant; \textit{$\Theta $} is the Debye temperature; \textit{D} and \textit{$\mu $} are the Kondo constants.

The fit of the resistivity data by formula \eqref{resist} gives good description of the low-temperature behavior but demonstrates strong (up to 15 \%) divergence from the experimental curves above 50 K (see Fig.~2).

At elevated temperatures, the major contribution in \eqref{resist} is given by second term $\rho _{\BG} (T)$ which has leading asymptotic $\rho _{\BG} (T)\propto AT/4\Theta ^{2} $ at $T>>\Theta $ and so cannot explain significant nonlinearity of $\rho (T)$. Thereby the standard electron-phonon model is not sufficient to explain the resistivity behavior at $T>T_{K} $.

It is known that the alloys and compounds containing transition metals demonstrate an additional contribution in electrical resistance due to Mott s--d interband scattering \cite{Mott}.  The contribution of that scattering in $\rho (T)$ is usually described by additional Mott term which has the form $\rho _{M} (T)=-\alpha T^{3} $, where $\alpha $ is the Mott constant. But, the $\rho _{M} (T)$ being added in the equation \eqref{resist} also does not describe the experimental data satisfactory over the whole temperature range (see Fig.~2). It is quite expectable because the Mott scattering should be small in Cu-Zr. Indeed, the magnetic properties of Cu-Zr alloys are determined by conductivity electrons only \cite{Glimois}, i.e. atoms have no localized magnetic moments and its $d$-level is completely filled.

 Note, that the Bloch--Gr\"{u}neisen model is based on the assumptions that the lattice vibrations are quasi-harmonic, and the Debye characteristic temperature is constant. It has been shown in a large number of reports \cite{Tosi,Flubacher} that the Debye temperatures for many substances have noticeable temperature dependence. Furthermore, these temperature variations of the Debye parameter also take place at cryogenic temperatures. Hence, the Debye temperature may be taken as a temperature function $\Theta (T)$ to describe the resistivity data properly. For simplicity, let us consider the linear function:
\begin{equation} \label{debye_temp}
\Theta (T)=\Theta _{0} +\alpha T,
\end{equation}
 where $\Theta $${}_{0}$ is the Debye temperature at the absolute zero, and $\alpha $ is the thermal coefficient.

 Using the relation \eqref{resist} with modification given by \eqref{debye_temp}, we have achieved excellent fitting of experimental data (see Fig.~2). The divergence between theoretical and experimental curves does not exceed 1.5 \%. The coefficients providing the best fits are listed in the Tab.~2.

 In order to illustrate full information on the Debye temperature behavior we plotted the concentration dependencies of $\Theta $ at different temperatures, see Fig.~3. It is clearly seen in the figure that the isothermal curves are significantly non-monotonic except the case of \textit{T}=0. This may be explained by strong structural distinctions of the investigated alloys. In particular, according to XRD analysis, each sample is characterized by its individual phase compositions including non-equilibrium crystal structures. In this case, it is unlikely to expect monotonic dependences of properties.

In order to get additional information about electronic structures of the alloys under consideration, we perform room temperature measurements of Hall constant $R_{H} $ and thermal conductivity $\lambda $. Using the relations $R_{H} =\mu \sigma =1/en_{e} $,  Wiedemann--Franz law $\lambda _{e} /\sigma =L_{0} T$ and the suggestion that thermal conductivity can be represented as the sum of electron  and phonon  contribution ($\lambda =\lambda _{e} +\lambda _{\rm ph} $), we estimate the set of key electronic characteristics, such as: charge carrier density $n_{e} $ and their mobility $\mu $, electron/phonon contributions in the transport properties and $\lambda /\lambda _{e} =L/L_{0} $ which represent the validity parameter of Wiedemann--Franz law. These characteristics are represented in Tab.~3.

\begin{figure}
  \centering
  \includegraphics[width=\columnwidth]{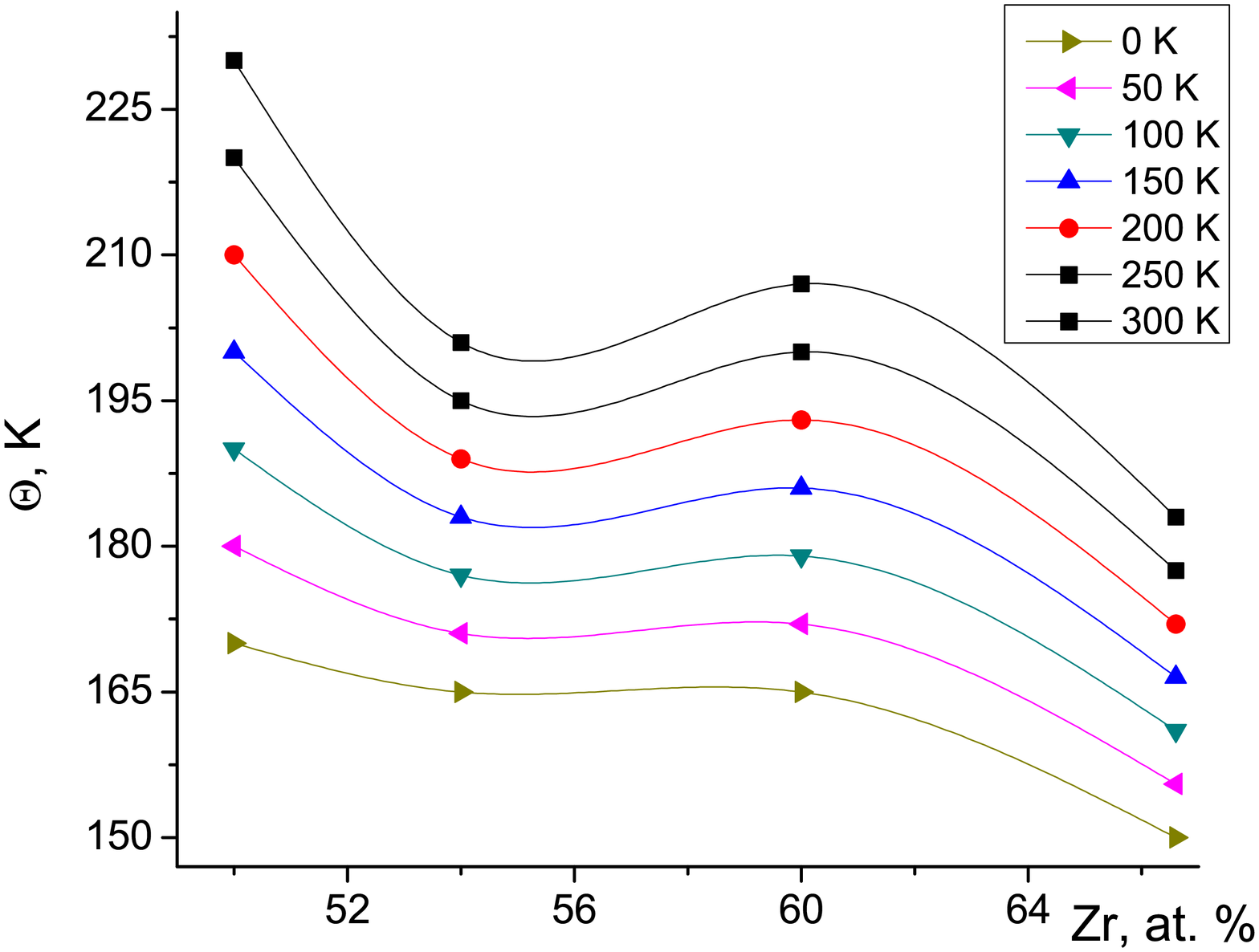}\\
  \caption{(Color online) The Debye characteristic temperatures of the Cu-Zr alloys at different compositions and temperatures.}
  \label{fig1:XRD}
\end{figure}

\begin{table*}
\centering
\scriptsize
\begin{tabular}{|c|c|c|c|c|} \hline
 & Cu${}_{50}$Zr${}_{50}$ & Cu${}_{46}$Zr${}_{54}$ & Cu${}_{40}$Zr${}_{60}$ & Cu${}_{33.4}$Zr${}_{66.6}$ \\ \hline
R${}_{H}$, 10${}^{11}$ m${}^{3}$/C & - 11.01 & - 6.57 & - 3.11 & - 4.07 \\ \hline
n${}_{e}$, 10${}^{28}$ m${}^{-3}$ & 5.7 & 9.5 & 20.1 & 15.4 \\ \hline
$\mu $, cm${}^{2}$/(V$\times$sec) & 1.11 & 0.86 & 0.59 & 1.15 \\ \hline
$\lambda $, W/(m$\times$K) & 9.10 & 11.6 & 22.0 & 27.8 \\ \hline
$\lambda $${}_{e}$, W/(m$\times$K) & 7.4 & 9.6 & 14.4 & 20.7 \\ \hline
$\lambda $${}_{\rm ph}$, W/(m$\times$K) & 1.7 & 2.0 & 7.6 & 7.1 \\ \hline
L(300 K)/L${}_{0}$ & 1.23 & 1.21 & 1.59 & 1.34 \\ \hline
\end{tabular}
\caption{The values of Hall coefficients, carrier density, electron mobility, thermal conductivity and the Lorenz relation of the Cu-Zr alloys at 300 K.}\label{Tab3}
\end{table*}

Hall coefficients are negative that indicates the charge carriers in these alloys are primarily electrons. The charge carrier density $n_{e}$ for the studied samples is similar to that in pure metals or conventional alloys. However, the Cu-Zr alloys have very low values of the charge carrier mobility $\mu $ which are comparable to those observed in the systems with strong lattice defects \cite{Kao}. In other words, free movement of carriers is confined due to the heavy structural defects (disorder). Thus, the lattice vibrations - phonons should play a noticeable role in the electron transport properties of the alloys. As seen in Tab.~3, the phonon contribution in all the samples is more than 20 \%, and its largest quantity has been obtained in Cu${}_{40}$Zr${}_{60}$ alloy. Recently, the similar behavior of the Lorenz relation and large phonon contribution has been observed in the glass-forming Al-based alloy \cite{Uporov}. It can be assumed that the observed features in Cu${}_{40}$Zr${}_{60}$ alloy are related with the existence of unusual monoclinic phase revealed by XRD analysis (see Fig.~1 and Tab.~1).

\section{Conclusions}

 A series of the Cu${}_{50-x}$Zr${}_{x}$ (x = 50, 54, 60 and 66.6) alloys were prepared using the standard arc-melting procedure. XRD analysis has revealed the existence of non-equilibrium martensitic structures in all the alloys except Cu${}_{33.4}$Zr${}_{66.6}$. For Cu${}_{40}$Zr${}_{60}$ and Cu${}_{46}$Zr${}_{54}$ alloys, the fraction of these phases is significant and, for Cu${}_{50}$Zr${}_{50}$, the only ones are presented. The martensitic monoclinic phase found in Cu${}_{40}$Zr${}_{60}$ alloy is rather unusual one: it has symmetry that differs from those usually observed for Cu-Zr system. These structural features are probably related with good glass-forming ability of the Cu-Zr system.

 Both the electrical resistivity values and Hall coefficients indicate the metallic conductance of the studied Cu-Zr samples and so electrons are the primary charge carries. At the temperatures lower than 20 K, the resistivity of the alloys exhibits Kondo-like behavior. This feature is probably related with small amounts of uncontrolled impurities of $d$-metals in the samples.

It was found that experimental temperature dependence of the resistivity cannot be described within framework of the standard Bloch--Gr\"{u}neisen model at elevated temperatures. We propose generalized  Bloch-Gr\"{u}neisen model taking into account the Debye temperature as a linear function. This improved relation describes experimental resistivity dependencies with high accuracy.

Because of the strong structure disorder in the Cu-Zr alloys, the carrier mobility estimated has very low values. The phonon contribution in thermal and electric conductivity is rather large that also indicates the strong structural distortions in the alloys. The largest deviations in the electron parameters, compared to other studied alloys, have been obtained for Cu${}_{40}$Zr${}_{60}$ sample. This behavior may be related with the existence of primitive monoclinic phase which is absent at other compositions.

\section{Acknowledgements}

This work was supported by Russian Scientific Foundation (grant RNF ¹14-13-00676). We would like to express special thanks to Dr. V. Bykov for the LFA measurements and valuable remarks.


\begin{thebibliography}{99}

\bibitem{Xu} D.H. Xu, B. Lohwongwatana, G. Duan, W.L. Johnson, C. Garland, Acta Mater. 52 (2004) 2621.

\bibitem{Wang1}  D. Wang, Li. Y, B.B. Sun, M.L. Sui, K. Lu, E. Ma, Appl. Phys. Lett. 84 (2004) 4029.

\bibitem{Wang2} W.H. Wang, J.J. Lewandowski, A.L. Greer, J. Mater. Res. 20 (2005) 2307.

\bibitem{Wang3} D. Wang, H. Tan, and Y. Li, Acta Mater. 53 (2005) 2969.

\bibitem{Shen} Y. Shen, E. Ma, and J. Xu, J. Mater. Sci. Technol. 24 (2008) 149.

\bibitem{Wang4}D. Wang, Y. Li, B. B. Sun, M. L. Sui, K. Lu, and E. Ma, Appl. Phys. Lett. 84 (2004) 4029.

\bibitem{Yang1} L. Yang, G.Q. Guo, L.Y. Chen et al., Phys. Rev. Lett. 109 (2012) 105502.

\bibitem{Yu} C.Y. Yu, X.J. Liu, J.Lu, G.P. Zheng and C.T. Liu, Sci. Rep. 3 (2013) 2124.

\bibitem{Bendert} J.C. Bendert, A.K. Gangopadhyay, N.A. Mauro and K. F. Kelton, Phys. Rev. Lett. 109 (2012) 185901.

\bibitem{Yang2} W. Yang, F. Liu, H. Liu, H.F. Wang, Z. Chen, G.C. Yang, J. All. Comp. 484 (2009) 702.

\bibitem{Kwon} Oh-Jib Kwon, Young-Kook Lee, Sang-Ok Park et al., Mat. Sci. Eng. A 449--451 (2007) 169.

\bibitem{Gantmakher1} V.F. Gantmakher and G.I. Kulesko, Sol. State Comm. 53 (1985) 267.

\bibitem{Gantmakher2} V.F. Gantmakher, G.I. Kulesko and Yu.B. Levin, JETP Lett. 39 (1984) 498.

\bibitem{Carvalho} E.M. Carvalho, I.R. Harris, J. Mat. Sci. 15 (1980) 1224.

\bibitem{Garoche} P. Garoche, J. Bigot, Phys. Rev. B 28 (1983) 6886.

\bibitem{Glimois} 16. J.~L. Glimois, P. Forey and J.L. Feron, J. Less-Comm. Met. 113 (1985) 213.

\bibitem{Fornell} J. Fornell, M.D. Baro, S. Surinach, A. Gebert and J. Sort, Adv. Eng. Mat. 13(1-2) (2011) 57.

\bibitem{Mott} N.F. Mott, H. Jones, The Theory of the Properties of Metals and Alloys, Oxford University Press, London, 1958.

\bibitem{Hasegawa} M. Hasegawa, H. Satob, T. Takeuchi, K. Soda, U. Mizutani, J. All. Comp. 483 (2009) 638.

\bibitem{Du} Jinglian Du, Bin Wen, Roderick Melnik, Yoshiyuki Kawazoe, J. All. Comp. 588 (2014) 96.

\bibitem{Chebotarev} Ya. Chebotarev, K.Avertsev, A.Golubev, M.Poddyakov X-Ray structure tabular processor RTP. 2002.\textbf{ }Copyright: freeware

\bibitem{Schryvers} D. Schryvers, G.S. Firstov, J.W. Seo, J, Van Humbeeck, Yu.N. Koval, Scripta Mat., 36 (1997) 119.

\bibitem{Okamoto} H. Okamoto, J. Phase Eq. Diff.33 (2012) 417.

\bibitem{Zhalko-Titarenko} A.V. Zhalko-Titarenko, M.L. Yevlashina, V.N. Antonov et al. Phys. Stat. Sol. (b). 184 (1994) 121.

\bibitem{Bauccio_book} ASM Metals Reference Book, 3rd Edition by Michael Bauccio, ASM International, Materials Park, 1993.

\bibitem{Abrikosov} A.A. Abrikosov, Fundamentals of the theory of metals, Amsterdam: North-Holland, 1988.

\bibitem{Kondo} Jun Kondo, Prog. Theor. Phys. 32 (1964) 37.

\bibitem{Tosi} M.P. Tosi and F.G. Fumi, Phys. Rev. 131 (1963) 1458.

\bibitem{Flubacher} P. Flubacher, A.~J. Leadbetter, and J.~A. Morrison, J. Chem. Phys. 33, (1960) 1751.

\bibitem{Kao} Y.-F. Kao et al., J. All. Comp. 509 (2011) 1607.

\bibitem{Uporov} S.~A. Uporov, V.~A. Bykov, D.~A. Yagodin, J. All. Comp. 589 (2014) 420.

\end{thebibliography}
\end{document}